\documentclass[12pt]{article}

\usepackage{scicite}
\usepackage{graphicx}
\usepackage{caption}
\usepackage{subfigure}
\usepackage{pdfpages}

\usepackage{times}

\newenvironment{sciabstract}{%
\begin{quote} \bf}
{\end{quote}}

\newcounter{lastnote}
\newenvironment{scilastnote}{%
\setcounter{lastnote}{\value{enumiv}}%
\addtocounter{lastnote}{+1}%
\begin{list}%
{\arabic{lastnote}.}
{\setlength{\leftmargin}{.22in}}
{\setlength{\labelsep}{.5em}}}
{\end{list}}

\title{Formation of matter-wave soliton trains by modulational instability}

\author
{Jason H.V. Nguyen, De Luo, and Randall G. Hulet$^{\ast}$\\
\\
\normalsize{Department of Physics and Astronomy, Rice University, Houston TX, 77005, USA}\\
\\
\normalsize{$^\ast$To whom correspondence should be addressed; E-mail:  randy@rice.edu.}
}

\date{}

\begin{document} 


\baselineskip24pt


\maketitle


\begin{sciabstract}
  Nonlinear systems can exhibit a rich set of dynamics that are inherently sensitive to their initial conditions.  One such example is modulational instability, which is believed to be one of the most prevalent instabilities in nature.  By exploiting a shallow zero-crossing of a Feshbach resonance, we characterize modulational instability and its role in the formation of matter-wave soliton trains from a Bose-Einstein condensate.  We examine the universal scaling laws exhibited by the system and, through real-time imaging, address a long-standing question of whether the solitons in trains are created with effectively repulsive nearest-neighbor interactions or rather evolve into such a structure.
\end{sciabstract}


Modulational instability (MI) is a process in which broadband perturbations spontaneously seed the nonlinear growth of a nearly monochromatic wave disturbance~\cite{Kamchatnov_Book}.  Owing to its generality, MI plays a role in a variety of different physical systems such as water waves, where it is known as a Benjamin-Feir instability~\cite{Benjamin_JFMech.27.417}; plasma waves; nonlinear optics~\cite{Stolli2012,Zakharov_PhysD.238.540,Hasegawa_Book}; and ultracold atomic gases~\cite{Kevrekidis_Mod.Phys.Lett.B.18.173}.  The nonlinear interaction resulting in MI also supports solitons, which are localized waves whose dispersion is exactly balanced by the nonlinearity~\cite{Zabusky_PhysRevLett.15.240,Zakharov_Sov.Phys.JETP.34.62}.  Thus, the rapid growth of fluctuations from MI, which leads to the breakup of the wave, is seen as a natural precursor to the formation of soliton trains.  In optical systems, this was first observed in the temporal domain\cite{Mollenauer.PhysRevLett.45.1095,Hasegawa_OptLett.9.288,Tai_PhysRevLett.56.135} and, subsequently, in the spatial domain\cite{Chen.RepProgPhys.75.8.2012}. 

Analogously, in an atomic Bose-Einstein condensate (BEC), MI drives the spontaneous formation of bright matter-wave solitons when the interaction between atoms is rapidly quenched from repulsive to attractive.  These systems are well described in most respects by the Gross-Pitaevskii equation, which is equivalent to the nonlinear Schr\"odinger equation with the addition of a harmonic trapping potential.    Here, the nonlinearity is determined by the s-wave scattering length, which is positive for a repulsive, defocusing nonlinearity and negative for an attractive, focusing one.  We will see that dissipation plays an important role in the matter-wave system, as it does in optical media.

Bright matter-wave solitons were first observed by applying an interaction ramp traversing a zero-crossing of the interaction parameter in a quasi-one-dimensional (quasi-1D) BEC~\cite{Khaykovich_Science.296.1290,Strecker_Nature.417.150}.  Several experiments have produced trains of up to $10$ solitons~\cite{Strecker_Nature.417.150,Cornish_PhysRevLett.96.170401}.  Because these solitons are harmonically confined they are not truly 1D and are susceptible to collapse resulting from the attractive nonlinearity.  This has the effect of limiting the number of atoms a single soliton can stably support~\cite{Perez_PhysRevA.57.3837,Gammal_PhysRevA.64.055602,Parker_JPhysB.40.3127}.  Additionally, solitons themselves may interact, exhibiting an effectively attractive or repulsive force that, according to mean-field theory, can be ascribed to a relative phase between solitons of ${\Delta\phi \approx 0}$ or ${\Delta\phi \approx \pi}$, respectively~\cite{Gordon_OptLett.8.596}.  These phase-dependent interactions were first observed in optical solitons\cite{Mitschke_Opt.Lett.12.355,Aitchison_Opt.Lett.16.15}.  In the case of matter-wave solitons, the peak density increases for in-phase collisions (${\Delta\phi \approx 0}$) which can produce annihilations and mergers, whereas out-of-phase collisions (${\Delta\phi \approx \pi}$) are expected to be more stable against collapse\cite{Baizakov_PhysRevA.70.053613,Khaykovich_PhysRevA.74.023607,Parker_JPhysB.41.045303,Parker_PhysicaD.238.1456.2009}.  These effects have been observed experimentally\cite{Nguyen_NPhys.918.2014}.  Solitons created in trains were found to be surprisingly stable, persisting for many cycles of oscillation in a harmonic trap despite being near the threshold for collapse\cite{Strecker_Nature.417.150,Cornish_PhysRevLett.96.170401}.  From this observation, it was inferred that an alternating-phase ($0$-$\pi$-$0$) structure was present, protecting the structure against collapse\cite{Strecker_Nature.417.150,Cornish_PhysRevLett.96.170401}. Detailed theoretical investigations have studied the formation of matter-wave soliton trains and attempted to explain the origins of the observed repulsive interaction between neighboring solitons~\cite{Khawaja_PhysRevLett.89.200404,Salasnich_PhysRevA.66.043603,Salasnich_PhysRevLett.91.080405,Brand_PhysRevLett.92.040401,Brand_PhysRevA.70.033607,Parker_PhysicaD.238.1456.2009,Davis_NJP.11.053017,Streltsov_PhysRevLett.106.240401}.  We address these issues in the experiments described here.

For MI, there is a positive feedback-driven exponential growth that is largest for the wavenumber $k_{MI}=\sqrt{4|a_f|n_{1D}} / a_{r} $\cite{Salasnich_PhysRevLett.91.080405,Brand_PhysRevLett.92.040401}.  Here, $a_{r}=\sqrt{\hbar / (m\omega_r})$ is the characteristic confinement length in the radial direction, $\hbar$ is Planck's constant divided by $2\pi$, $m$ is the atomic mass, $\omega_r$ is the radial frequency of the harmonic trap, $a_f$ is the (negative) scattering length after the quench, and $n_{1D}$ is the line density of the condensate before the quench.  The healing length, $\xi=k_{MI}^{-1}$, naturally lends itself as the characteristic length scale for MI in this system; correspondingly, the rate at which fluctuations grow sets a characteristic time scale given by $\gamma^{-1}$, where $\gamma=\hbar k_{MI}^2 / (2 m)$.  

Once the scattering length is quenched from positive to negative, the effects of MI manifest as density modulations of the gas (Fig.~$1$A).  The atoms first clump together into regions of increased density, owing to the nonlinear focusing of the attractive interaction. Regions of high density, separated by a spatial distance of $2\pi\xi$ appear on a timescale given by $\gamma^{-1}$.  These density clumps evolve into solitons whose dispersion is balanced by the nonlinear attraction between atoms.

Although it is clear that MI is crucial to the formation of matter-wave soliton trains~\cite{Khawaja_PhysRevLett.89.200404,Salasnich_PhysRevA.66.043603,Salasnich_PhysRevLett.91.080405,Brand_PhysRevLett.92.040401,Brand_PhysRevA.70.033607,Davis_NJP.11.053017}, the identification of the mechanism responsible for their stability has remained elusive.  Several theories have been proposed.  In the simulations of Ref.~\citen{Khawaja_PhysRevLett.89.200404}, the authors determined the spectrum of the phase of the wavefunction produced by quantum fluctuations when the scattering length was suddenly quenched.  They imprinted the condensate wavefunction with this phase and found the subsequent development of an alternating phase structure and dynamics that match those of the experiment~\cite{Strecker_Nature.417.150}.

In another study~\cite{Salasnich_PhysRevA.66.043603}, similar dynamics were calculated with the use of an effective time-dependent 1D nonpolynomial Schr\"odinger equation, but an alternating phase structure was simply imprinted onto the solitons.  In a subsequent paper~\cite{Salasnich_PhysRevLett.91.080405}, imprinting the condensate with an ad hoc phase structure was shown to be unnecessary; a nearly alternating-phase structure emerged in numerical simulations by allowing the phase of the condensate to evolve self-consistently according to a Gross-Pitaevskii equation that included a dissipative three-body term.  

In Ref.~\citen{Brand_PhysRevLett.92.040401}, self-interference, rather than quantum fluctuations served to seed MI.  Exponential growth of these fringes first led to primary collapse in cases where the atom number of an individual soliton exceeded a critical value during the early part of MI.  The resultant solitons in the train were found to have arbitrary phases.  To acquire an alternating-phase structure,  it was proposed that a stage of secondary collapses occurred, wherein binary collisions between solitons resulted in annihilations and mergers of near in-phase soliton pairs.  These collisions would serve to distill the soliton train, resulting in the eventual formation of alternating phases but accompanied by the loss of a large number of atoms\cite{Brand_PhysRevLett.92.040401}.   

In a subsequent comparison between MI seeded by noise and self-interference~\cite{Brand_PhysRevA.70.033607}, it was determined that both should contribute to MI at comparable timescales.  By varying the noise added into their simulations, the authors were able to identify regimes dominated by self-interference or noise.  Notably, with MI seeded by self-interference, soliton formation occurred at the edges of the condensate first, since the fringes from self-interference first achieve a sufficiently long wavelength there~\cite{Brand_PhysRevLett.92.040401,Brand_PhysRevA.70.033607}.  In contrast, MI seeded by noise led to the development of solitons first in the center, where the density and the rate of MI was highest, and finally at the edges~\cite{Brand_PhysRevA.70.033607}.

In the mean-field approaches discussed thus far, the observed stability against secondary collapses is attributed to an alternating-phase structure, although whether this alternating-phase structure is initially present, or evolves out of the mutual annihilation of attractively interacting solitons has been debated~\cite{Khawaja_PhysRevLett.89.200404,Salasnich_PhysRevA.66.043603,Salasnich_PhysRevLett.91.080405,Brand_PhysRevLett.92.040401,Brand_PhysRevA.70.033607}.  Approaches extending beyond mean-field theory, such as the truncated Wigner approximation (TWA) in $1$D~\cite{Davis_NJP.11.053017} or the  multiconfigurational time-dependent Hartree for bosons (MCTDHB) method~\cite{Streltsov_PhysRevLett.100.130401,Streltsov_PhysRevLett.106.240401,Sakmann_NPhys.12.451.2016} suggest that quantum effects may produce effectively repulsive interactions, independent of the relative phase.  The extension of the TWA to $3$D~\cite{Davis_NJP.11.053017}, however, resulted in a rapid loss of solitons that contradicts observations~\cite{Strecker_Nature.417.150,Cornish_PhysRevLett.96.170401}.  Furthermore, convergence with the MCTDHB method has been shown to be pathological for bosons with attractive interactions, which may have affected previous conclusions~\cite{Cosme_PhysRevA.94.043603}.

We address these issues with a degenerate gas of $^{7}\mathrm{Li}$ atoms.  Our methods have been described elsewhere~\cite{SOM_ref,Nguyen_NPhys.918.2014}.  A BEC of atoms in the $|F=1,m_F=1\rangle$ state (where $F$ and $m_F$ are the quantum numbers of the total atomic angular momentum and its projection, respectively) is confined in a cylindrically symmetric harmonic trap with radial and axial oscillation frequencies of $\omega_r/ 2\pi= 346\ \mathrm{Hz}$ and $\omega_z/ 2\pi= 7.4\ \mathrm{Hz}$, respectively.  The interaction between atoms is magnetically controlled via a broadly tunable Feshbach resonance~\cite{Pollack_PhysRevLett.102.090402,Dyke_PhysRevA.88.023625}, and is initially set to a scattering length of $a_i\approx 3\ a_0$ (where $a_0$ is the Bohr radius)~\cite{SOM_ref}.  We quench the interaction to a final scattering length, $a_f<0$, in a linear ramp time of $t_r=1\ \mathrm{ms}$.  After waiting a variable hold time, $t_h$, we take an \emph{in situ} polarization phase-contrast image (PPCI)~\cite{Bradley_PhysRevLett.78.985}.  Our PPCI method can be minimally destructive, resulting in the loss of $<2\%$ of atoms per image, thus enabling a sequence of images of the same soliton train.  

The formation of a soliton train is shown in Fig.~$1$B for a scattering length of $a_f=-0.18\ a_0$, with $t_h$ from $0$ to $20\ \mathrm{ms}$ (each of these images corresponds to a different experimental run).  The images in Fig.~$1$C depict the formation for a scattering length of $a_f=-2.5\ a_0$, highlighting some key differences between smaller and larger $|a_f|$. For larger $|a_f|$ we find that the formation occurs on a faster timescale and we also see a reduction in the number of solitons remaining with increasing $t_h$. We characterize the effect of MI on the density profile of the BEC by defining a contrast parameter, $\eta$, which is a measure of the deviation in the density from a Thomas-Fermi profile~\cite{SOM_ref}.   We observe rapid growth of $\eta$ in the central region as compared with the sides of the condensate~\cite{SOM_ref}.  According to Ref.~\citen{Brand_PhysRevA.70.033607}, this implies that the seed for MI is dominated by noise, which may be technical, thermal, or quantum in origin, rather than self-interference.

The loss of total atom number, $N_a$ vs. $t_h$, is plotted in Fig.~$2$A.  We observe an initial plateau where $N_a$ changes little, followed by a period of rapid atom loss.  The plateau and subsequent atom loss are reminiscent of experiments exploring the collapse of an attractive condensate of $^{85}\mathrm{Rb}$ atoms~\cite{Donley_Nature.412.295,Kagan_PhysRevA.84.033632}.  MI provides a simple and intuitive explanation for this initial plateau.  When $t_r$ is fast compared with $\gamma^{-1}$, the dynamics are initially frozen out.  This timescale is indicated by the arrows in Fig.~$2$A, calculated for several values of $a_f$.  As $|a_f|$ is increased, $\gamma^{-1}$ is predicted to become smaller, in agreement with the data.  Solitons are formed for times longer than $\gamma^{-1}$ .

The universality of the MI timescale and of atom loss becomes evident when $t_h$ is rescaled by $\gamma^{-1}$ (Fig. $2$B).  We find that the data collapse onto a single curve, with the exception of ${a_f=-2.5\ a_0}$.  Because ${t_r = 1\ \mathrm{ms}> \gamma^{-1}=0.42\ \mathrm{ms}}$ for this scattering length, the plateau is notably absent.  For all other scattering lengths, the onset of atom loss begins shortly after $t_h \gamma = 1$.  We fit the data (Fig.~$2$B) for $t_h > \gamma^{-1}$ to a power law decay, where $N_a=N_0(\gamma t_h)^\kappa$ with $\kappa=-0.35(1)$ (Here $N_0$ is the total initial number of atoms and $\kappa$ is the power law exponent).

Scaling laws within the system also provide us with a simple yet surprisingly accurate estimate of the number of solitons, $N_s$, formed by MI.  Assuming an initial condensate length of $2R_{TF}$, where $R_{TF}$ is the Thomas-Fermi radius, we estimate $N_s\simeq 2R_{TF} / (2\pi\xi)$ from simple length scale arguments~\cite{Khawaja_PhysRevLett.89.200404,Salasnich_PhysRevLett.91.080405}.  Because the dynamics of the system are frozen for fast $t_r$ (as compared with $\gamma^{-1}$), the initial conditions are entirely determined by $|a_f|$.  MI produces a modulation of the density, with the density of defects set by $(2\pi\xi)^{-1}$.  In our experiments, $R_{TF}$ is held constant, whereas $\xi$ is controlled by changing $a_f$.  In Fig.~$3$A we plot $N_s$ vs. $a_f$, and find excellent agreement with this simple model for ${|a_f|<1\ a_0}$.  For larger $|a_f|$, $N_s$ is limited by primary collapses that arise when the number of atoms for a single soliton exceeds the critical number for collapse, $N_c=0.67a_{r} / |a_f|$, where the factor of $0.67$ accounts for the aspect ratio of the trapping potential\cite{Perez_PhysRevA.57.3837,Gammal_PhysRevA.64.055602,Parker_JPhysB.40.3127}.  Furthermore, solitons are able to undergo primary collapse during the quench for $t_r > \gamma^{-1}$.

To examine whether primary collapses, or secondary collapses that arise from annihilations or mergers contribute to the observed decrease in $N_a$, we plot $N_s$ vs. $t_h$ in Fig.~$3$B. We find that for the two smallest $|a_f|$, $a_f = -0.18\ a_0$ and $-0.42\ a_0$, $N_s$ remains constant with increasing $t_h$, indicating that neither primary nor secondary collapses have occurred. The fact that $N_s$ remains constant indicates that the interactions between neighboring solitons are dominantly repulsive, thus suppressing secondary collapses. $N_s$ decreases for larger values of $|a_f|$, indicating the effect of collapse.  Because the collisional time scale is expected to be on the order of the breathing mode period, $t_{br} = 68\ \mathrm{ms}$, we attribute the initial rapid $(t_h < 20\ \mathrm{ms})$ soliton loss to primary collapses.  Secondary collapses are likely to play a role at later times, particularly for the $a_f = -2.5\ a_0$ data, for which soliton loss is observed until $N_s \approx 2$.  Additional insight into the appearance of collapse may be obtained by examining the strength of the nonlinearity, $\Delta$, which we define as the number of atoms per soliton, normalized to the critical number, $\Delta = N_a/(N_s N_c)$ (Fig.~$3$C). For both $a_f = -0.18\ a_0$ and $-0.42\ a_0$ the initial $\Delta < 0.6$, and $\Delta$ decays only because of the loss of atoms from each independent soliton, not by losing solitons. On the other hand, the large initial value of $\Delta$ for larger $|a_f|$ explains the relative instability to collapse exhibited by these solitons.

To gain further insight into the nature of atom loss, we fit the decay in $N_a$ to a function that assumes that atoms are lost from independent solitons by three-body recombination (Fig.~S3).  This assumption yields a power law decay, as observed, but with $\kappa = -0.5$ or $-0.25$, depending on assumptions regarding the soliton length~\cite{SOM_ref}.  These values bracket the measured exponent of $-0.35$.  We extract a three-body loss coefficient, $L_3$, from this analysis and find that it ranges between $10^{-26}$ and $10^{-25}\ \mathrm{cm^6 /s}$, depending on the initial assumptions~\cite{SOM_ref}.  This is much greater than values previously measured for small positive scattering lengths of $10^{-28}\ \mathrm{cm^6 /s}$ ~\cite{Dyke_PhysRevA.88.023625}.  Additionally, when the scattering length is ramped slowly ($>250\ \mathrm{ms}$) rather than suddenly quenched, the loss rate is below our ability to measure ($L_3 < 10^{-28}\ \mathrm{cm^6 /s}$) (Fig.~S3).  We conclude that the much larger rate of loss arises from dynamical changes in the density induced by the sudden quench.  One consequence is the excitation of a breathing mode that periodically modulates the density, and thus the rate of three-body loss.  The loss rate plateau seen for $t_h > t_{br} / 2$ in Fig.~$2$A is likely a manifestation of this effect.  The quench may also induce partial collapses that originate in localized high-density regions of a soliton.  The resulting atom loss can self-arrest the collapse, thus resulting in a series of intermittent, partial collapses~\cite{Kagan_PhysRevLett.81.933,Saito_PhysRevLett.86.1406}.

Our minimally-destructive imaging technique allows us to take multiple images of the same soliton train to directly observe the dynamics.  These images for the small $|a_f|$ data confirm the expected repulsive soliton-soliton interactions.  Two such examples are shown for $a_f=-0.18\ a_0$ in Figs.~$4$, A and B.  We find that the solitons remain well-separated from one another at \emph{all} times, from which we infer dominantly repulsive interactions, even as the soliton train first emerges.

We have examined MI in detail, elucidating its universal role in the spontaneous formation of matter-wave soliton trains.  Our results indicate that MI in this context is driven by noise, and that for small $|a_f|$, neighboring solitons already interact repulsively during the initial formation of the soliton train, independent of secondary collisions.  This may also be the case for larger $|a_f|$, but primary collapse dominates the dynamics in this case, and the soliton train quickly dissipates as a result.  Similar phase and wavelength correlations have been observed in optical MI experiments~\cite{Stolli2012}.  We have also demonstrated natural scaling laws for atom loss. The scaling behavior is similar to systems that are described by the Kibble-Zurek mechanism~\cite{Kibble_JPhysA.9.1387,Zurek_Nature.317.505,Polkovnikov_RevModPhys.83.863}, although a key difference in our system is the presence of dissipation and collapse, which is not part of the Kibble-Zurek scenario.  The ability to finely control the interaction between atoms, and the relatively slow timescale for dynamics point toward the study of rogue matter-waves\cite{Bludov_PhysRevA.80.033610,Wen_EPJD.64.473}, analogous to the rogue waves observed in optical systems~\cite{Solli_Nature.450.1054}, as a natural extension of this work.  Our methods are additionally amenable to studying the formation and propagation of higher-order solitons, such as breathers\cite{Kevrekidis_PhysRevLett.90.230401,Sakaguchi_PhysRevE.70.066613}.

\textbf{Note added in proof:} A manuscript reporting modulational instability in $^{85}\mathrm{Rb}$\cite{2017arXiv170307502E} was posted subsequent to the submission of this manuscript.

\bibliography{scibib}

\begin{thebibliography}{10}

\bibitem{Kamchatnov_Book}
A.~M. Kamchatnov, {\it Nonlinear periodic waves and their modulations\/} (World
  Scientific, 2000).

\bibitem{Benjamin_JFMech.27.417}
T.~B. Benjamin, J.~E. Feir, {\it J. Fluid Mech.\/} {\bf 27}, 417 (1967).

\bibitem{Stolli2012}
D.~R. Solli, G.~Herink, B.~Jalali, C.~Ropers, {\it Nat. Photon.\/} {\bf 6}, 463
  (2012).

\bibitem{Zakharov_PhysD.238.540}
V.~E. Zakharov, L.~A. Ostrovsky, {\it Physica D\/} {\bf 238}, 540  (2009).

\bibitem{Hasegawa_Book}
A.~Hasegawa, Y.~Kodama, {\it Solitons in optical communications\/} (Claredon
  Press, 1995).

\bibitem{Kevrekidis_Mod.Phys.Lett.B.18.173}
P.~G. Kevrekidis, D.~J. Frantzeskakis, {\it Mod. Phys. Lett. B\/} {\bf 18}, 173
  (2004).

\bibitem{Zabusky_PhysRevLett.15.240}
N.~J. Zabusky, M.~D. Kruskal, {\it Phys. Rev. Lett.\/} {\bf 15}, 240 (1965).

\bibitem{Zakharov_Sov.Phys.JETP.34.62}
V.~E. Zakharov, A.~B. Shabat, {\it Sov. Phys.--JETP\/} {\bf 34}, 62 (1972).

\bibitem{Mollenauer.PhysRevLett.45.1095}
L.~F. Mollenauer, R.~H. Stolen, J.~P. Gordon, {\it Phys. Rev. Lett.\/} {\bf
  45}, 1095 (1980).

\bibitem{Hasegawa_OptLett.9.288}
A.~Hasegawa, {\it Opt. Lett.\/} {\bf 9}, 288 (1984).

\bibitem{Tai_PhysRevLett.56.135}
K.~Tai, A.~Hasegawa, A.~Tomita, {\it Phys. Rev. Lett.\/} {\bf 56}, 135 (1986).

\bibitem{Chen.RepProgPhys.75.8.2012}
Z.~Chen, M.~Segev, D.~N. Christodoulides, {\it Rep. Prog. Phys.\/} {\bf 75},
  086401 (2012).

\bibitem{Khaykovich_Science.296.1290}
L.~Khaykovich {\it et~al.\/}, {\it Science\/} {\bf 296}, 1290 (2002).

\bibitem{Strecker_Nature.417.150}
K.~E. Strecker, G.~Partridge, A.~G. Truscott, R.~G. Hulet, {\it Nature\/} {\bf
  417}, 150 (2002).

\bibitem{Cornish_PhysRevLett.96.170401}
S.~L. Cornish, S.~T. Thompson, C.~E. Wieman, {\it Phys. Rev. Lett.\/} {\bf 96},
  170401 (2006).

\bibitem{Perez_PhysRevA.57.3837}
V.~M. P\'erez-Garc\'{\i}a, H.~Michinel, H.~Herrero, {\it Phys. Rev. A\/} {\bf
  57}, 3837 (1998).

\bibitem{Gammal_PhysRevA.64.055602}
A.~Gammal, T.~Frederico, L.~Tomio, {\it Phys. Rev. A\/} {\bf 64}, 055602
  (2001).

\bibitem{Parker_JPhysB.40.3127}
N.~G. Parker, S.~L. Cornish, C.~S. Adams, A.~M. Martin, {\it J. Phys. B: At.,
  Mol. Opt. Phys.\/} {\bf 40}, 3127 (2007).

\bibitem{Gordon_OptLett.8.596}
J.~P. Gordon, {\it Opt. Lett.\/} {\bf 8}, 596 (1983).

\bibitem{Mitschke_Opt.Lett.12.355}
F.~M. Mitschke, L.~F. Mollenauer, {\it Opt. Lett.\/} {\bf 12}, 355 (1987).

\bibitem{Aitchison_Opt.Lett.16.15}
J.~S. Aitchison {\it et~al.\/}, {\it Opt. Lett.\/} {\bf 16}, 15 (1991).

\bibitem{Baizakov_PhysRevA.70.053613}
B.~B. Baizakov, B.~A. Malomed, M.~Salerno, {\it Phys. Rev. A\/} {\bf 70},
  053613 (2004).

\bibitem{Khaykovich_PhysRevA.74.023607}
L.~Khaykovich, B.~A. Malomed, {\it Phys. Rev. A\/} {\bf 74}, 023607 (2006).

\bibitem{Parker_JPhysB.41.045303}
N.~G. Parker, A.~M. Martin, S.~L. Cornish, C.~S. Adams, {\it J. Phys. B: At.,
  Mol. Opt. Phys.\/} {\bf 41}, 045303 (2008).

\bibitem{Parker_PhysicaD.238.1456.2009}
N.~G. Parker, A.~M. Martin, C.~S. Adams, S.~L. Cornish, {\it Physica D\/} {\bf
  238}, 1456  (2009).

\bibitem{Nguyen_NPhys.918.2014}
J.~H.~V. Nguyen, P.~Dyke, D.~Luo, B.~A. Malomed, R.~G. Hulet, {\it Nat.
  Phys.\/} {\bf 10}, 918 (2014).

\bibitem{Khawaja_PhysRevLett.89.200404}
U.~Al~Khawaja, H.~T.~C. Stoof, R.~G. Hulet, K.~E. Strecker, G.~B. Partridge,
  {\it Phys. Rev. Lett.\/} {\bf 89}, 200404 (2002).

\bibitem{Salasnich_PhysRevA.66.043603}
L.~Salasnich, A.~Parola, L.~Reatto, {\it Phys. Rev. A\/} {\bf 66}, 043603
  (2002).

\bibitem{Salasnich_PhysRevLett.91.080405}
L.~Salasnich, A.~Parola, L.~Reatto, {\it Phys. Rev. Lett.\/} {\bf 91}, 080405
  (2003).

\bibitem{Brand_PhysRevLett.92.040401}
L.~D. Carr, J.~Brand, {\it Phys. Rev. Lett.\/} {\bf 92}, 040401 (2004).

\bibitem{Brand_PhysRevA.70.033607}
L.~D. Carr, J.~Brand, {\it Phys. Rev. A\/} {\bf 70}, 033607 (2004).

\bibitem{Davis_NJP.11.053017}
B.~J. D\c{a}browska-W\"{u}ster, S.~W\"{u}ster, M.~J. Davis, {\it New J.
  Phys.\/} {\bf 11}, 053017 (2009).

\bibitem{Streltsov_PhysRevLett.106.240401}
A.~I. Streltsov, O.~E. Alon, L.~S. Cederbaum, {\it Phys. Rev. Lett.\/} {\bf
  106}, 240401 (2011).

\bibitem{Streltsov_PhysRevLett.100.130401}
A.~I. Streltsov, O.~E. Alon, L.~S. Cederbaum, {\it Phys. Rev. Lett.\/} {\bf
  100}, 130401 (2008).

\bibitem{Sakmann_NPhys.12.451.2016}
K.~Sakmann, M.~Kasevich, {\it Nat. Phys.\/} {\bf 12}, 451 (2016).

\bibitem{Cosme_PhysRevA.94.043603}
J.~G. Cosme, C.~Weiss, J.~Brand, {\it Phys. Rev. A\/} {\bf 94}, 043603 (2016).

\bibitem{SOM_ref}
See the supplementary materials on Science Online.

\bibitem{Pollack_PhysRevLett.102.090402}
S.~E. Pollack {\it et~al.\/}, {\it Phys. Rev. Lett.\/} {\bf 102}, 090402
  (2009).

\bibitem{Dyke_PhysRevA.88.023625}
P.~Dyke, S.~E. Pollack, R.~G. Hulet, {\it Phys. Rev. A\/} {\bf 88}, 023625
  (2013).

\bibitem{Bradley_PhysRevLett.78.985}
C.~C. Bradley, C.~A. Sackett, R.~G. Hulet, {\it Phys. Rev. Lett.\/} {\bf 78},
  985 (1997).

\bibitem{Donley_Nature.412.295}
E.~A. Donley {\it et~al.\/}, {\it Nature\/} {\bf 412}, 295 (2001).

\bibitem{Kagan_PhysRevA.84.033632}
P.~A. Altin {\it et~al.\/}, {\it Phys. Rev. A\/} {\bf 84}, 033632 (2011).

\bibitem{Kagan_PhysRevLett.81.933}
{\relax Yu}.~Kagan, A.~E. Muryshev, G.~V. Shlyapnikov, {\it Phys. Rev. Lett.\/}
  {\bf 81}, 933 (1998).

\bibitem{Saito_PhysRevLett.86.1406}
H.~Saito, M.~Ueda, {\it Phys. Rev. Lett.\/} {\bf 86}, 1406 (2001).

\bibitem{Kibble_JPhysA.9.1387}
T.~W.~B. Kibble, {\it J. Phys. A--Math Gen.\/} {\bf 9}, 1387 (1976).

\bibitem{Zurek_Nature.317.505}
W.~H. Zurek, {\it Nature\/} {\bf 317}, 505 (1985).

\bibitem{Polkovnikov_RevModPhys.83.863}
A.~Polkovnikov, K.~Sengupta, A.~Silva, M.~Vengalattore, {\it Rev. Mod. Phys.\/}
  {\bf 83}, 863 (2011).

\bibitem{Bludov_PhysRevA.80.033610}
{\relax Yu}.~V. Bludov, V.~V. Konotop, N.~Akhmediev, {\it Phys. Rev. A\/} {\bf
  80}, 033610 (2009).

\bibitem{Wen_EPJD.64.473}
L.~Wen {\it et~al.\/}, {\it Eur. Phys. J. D\/} {\bf 64}, 473 (2011).

\bibitem{Solli_Nature.450.1054}
D.~R. Solli, C.~Ropers, P.~Koonath, B.~Jalali, {\it Nature\/} {\bf 450}, 1054
  (2007).

\bibitem{Kevrekidis_PhysRevLett.90.230401}
P.~G. Kevrekidis, G.~Theocharis, D.~J. Frantzeskakis, B.~A. Malomed, {\it Phys.
  Rev. Lett.\/} {\bf 90}, 230401 (2003).

\bibitem{Sakaguchi_PhysRevE.70.066613}
H.~Sakaguchi, B.~A. Malomed, {\it Phys. Rev. E\/} {\bf 70}, 066613 (2004).

\bibitem{2017arXiv170307502E}
P.~J. {Everitt} {\it et~al.\/}, {\it arXiv:1703.07502\/}  (2017).

\end{thebibliography}

\bibliographystyle{Science}


\begin{scilastnote}
\item We thank K. Hazzard, L. Carr, E. Mueller, and B. Malomed for helpful discussions. This work was supported by the NSF (Grants PHY-$1408309$ and PHY-$1607215$), the Welch Foundation (Grant No. C-$1133$), an ARO-MURI (Grant No. W911NF-14-1-0003) and the ONR.
\end{scilastnote}


\clearpage

\begin{figure}
	\centering
\includegraphics[width=1.00\textwidth]{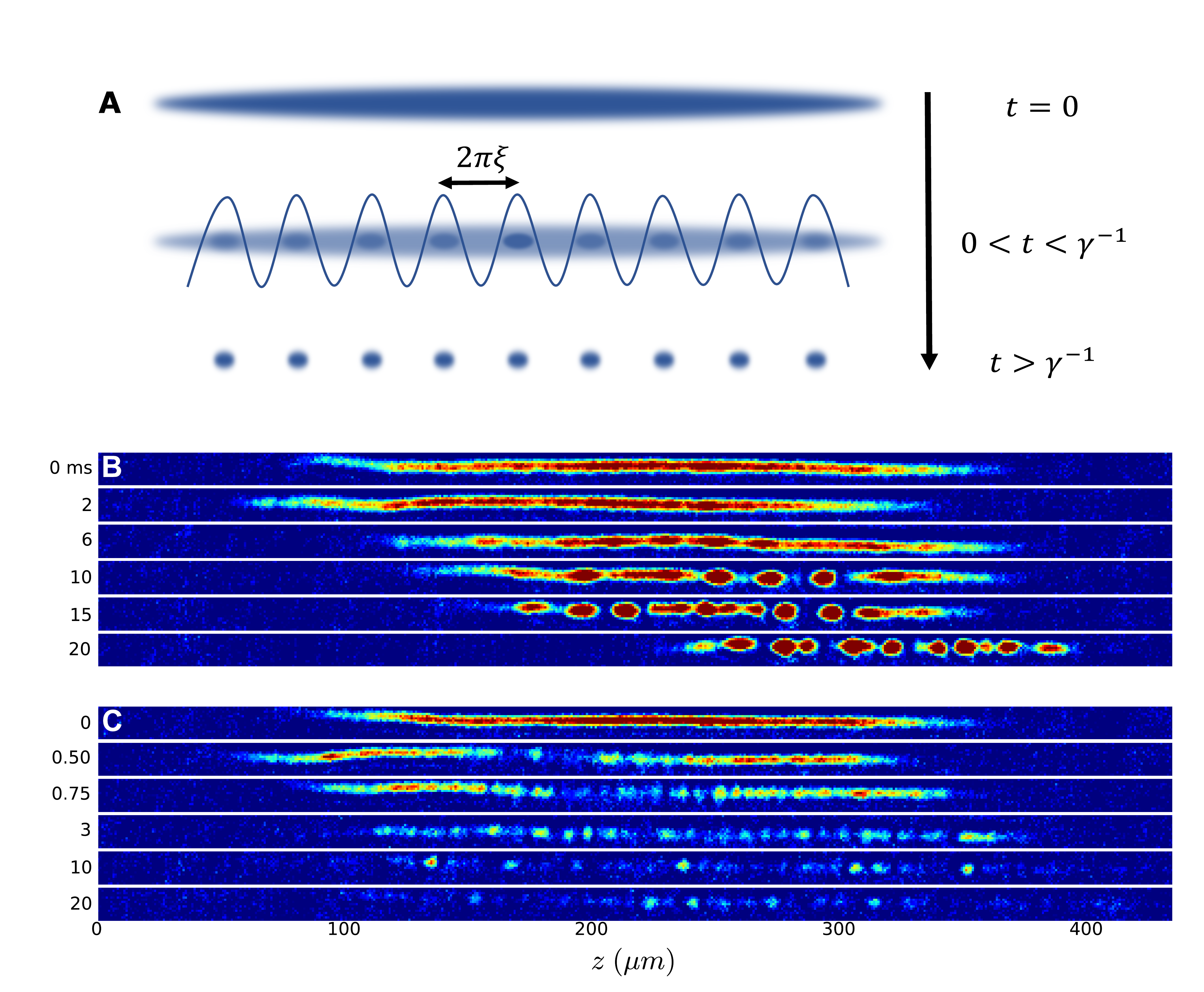}
	\caption*{\textbf{Fig.~1. Soliton train formation from modulational instability (MI).}  \textbf{(A)} Schematic representation of the effects of a scattering length quench.  At short times, the condensate has not responded to changes in scattering length.  MI results in rapid growth of fluctuations at a length scale of $2\pi\xi$.  Atoms flow towards regions of high density on a timescale of $\gamma^{-1}$ owing to a nonlinear focusing from attractive interactions.  Solitons are formed for $t>\gamma^{-1}$.  \textbf{(B)} Column density images for $a_f=-0.18\ a_0$. Immediately after the quench, there is no discernible change in $N_a$, nor is there any change in the shape from that of the original condensate at $a_i=+3\ a_0$.  Solitons form at later times, and undergo breathing and dipole oscillations.  \textbf{(C)} Similar to \textbf{B}, except with $a_f=-2.5\ a_0$.  Modulations appear much earlier, as do gaps near the center where the density of the original condensate was high, which we attribute to primary collapses.  A reduction in $N_s$ is evident at longer $t_h$.  Each image corresponds to a different experimental run, and hence real-time dynamics cannot be directly inferred from these images.  Here, $z$ is the position along the axial coordinate.}
	\label{fig:fig1}
\end{figure}

\begin{figure}
	\centering
		\includegraphics[width=1.00\textwidth]{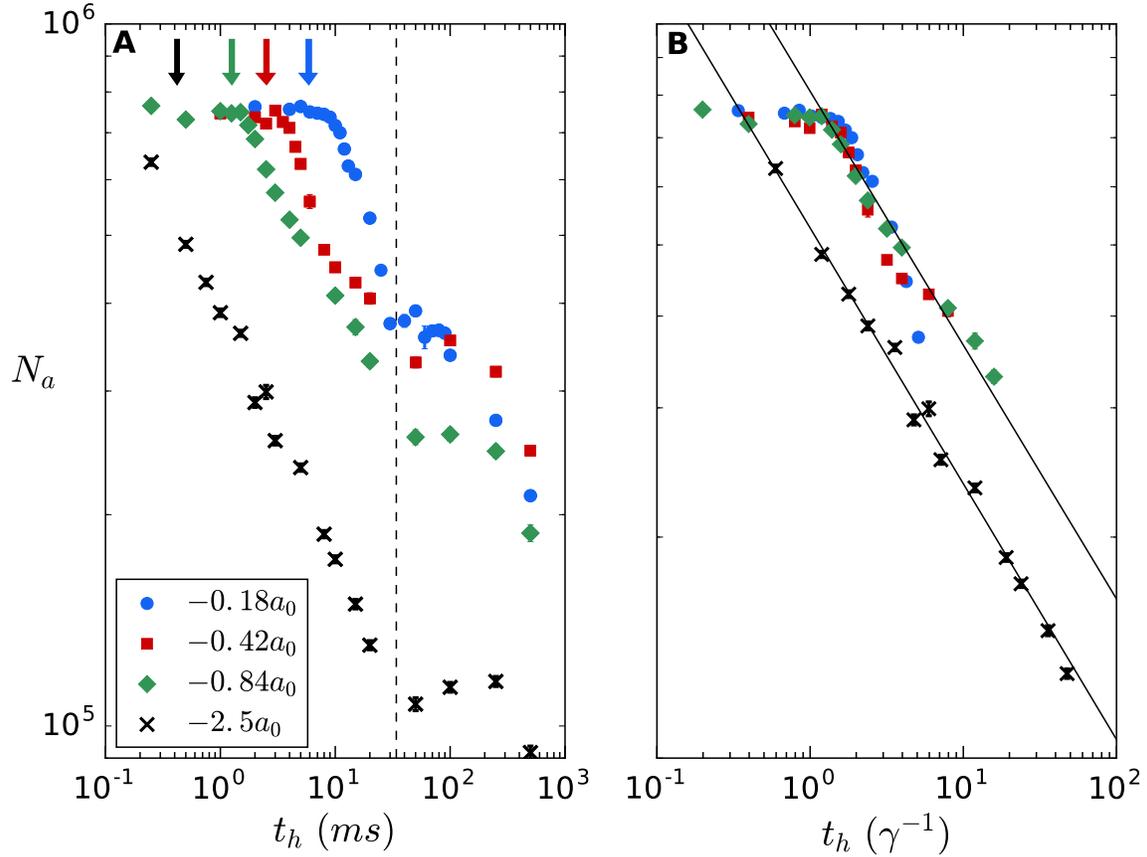}
	\caption*{\textbf{Fig.~2. Post-quench evolution of atom number.} \textbf{(A)}  $N_a$ vs. $t_h$ for various $a_f$.  The arrows indicate the calculated $\gamma^{-1}$ for each value of $a_f$, which is determined using the peak value of $n_{1D}$.  The black dashed line corresponds to half of a breathing period ($t_{br}=68\ \mathrm{ms}$).  We observe a plateau in $N_a$ for each $a_f$, followed by a rapid decrease in atoms starting shortly after $t_h\approx \gamma^{-1}$.  We attribute the lack of a plateau for $a_f=-2.5\ a_0$ to $t_r > \gamma^{-1}$.  \textbf{(B)}  Data replotted vs. $t_h \gamma$.  The data collapse onto a single curve, except for $a_f=-2.5\ a_0$.  The data are fit to a power law, $N_a=N_0 (t_h \gamma)^{\kappa}$, shown as a solid black line, where $\kappa=-0.35(1)$ for both fits.  For all $a_f$, points for $t_h> t_{br} /2$ have been omitted from the fit.  Error bars are the standard deviation of the mean of up to $30$ shots.}
	\label{fig:fig2}
\end{figure}

\begin{figure}
	\centering
		\includegraphics[width=1.00\textwidth]{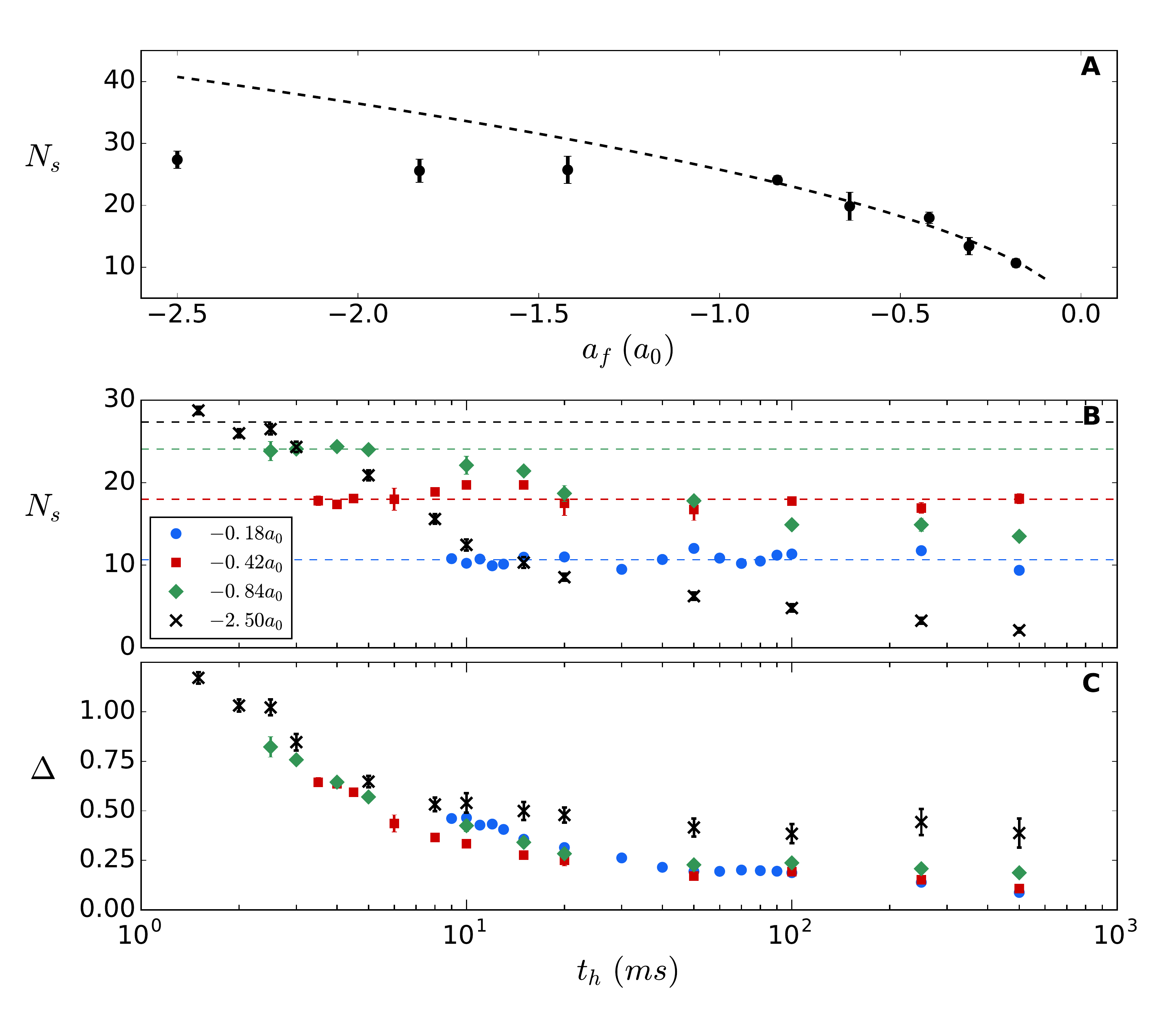}
	\caption*{\textbf{Fig.~3. Post-quench evolution of soliton number and strength of nonlinearity.}  \textbf{(A)} $N_s$ vs. $a_f$.  The dashed line corresponds to a fit of the data to the model (see text), where an overall scaling of $1.04(2)$ is the only fit parameter.  Data for $|a_f|>1\ a_0$ are omitted from the fit.  We attribute the suppression in $N_s$ for $|a_f|>1\ a_0$ to primary collapse, resulting in a reduction in the number of solitons formed. \textbf{(B)} $N_s$ vs. $t_h$.  $N_s$ does not change with $t_h$ for the two smallest $|a_f|$, whereas for larger $|a_f|$,  $N_s$ decays with $t_h$.  Dashed lines correspond to the initial number of solitons.  \textbf{(C)} $\Delta$ vs. $t_h$.  The initial value of $\Delta=N_a / (N_s N_c)$ increases as $|a_f|$ is increased, and is consistent with an expected $\sqrt{|a_f|}$ scaling.  This trend continues up to $\Delta=1$, above which the solitons are unstable against primary collapse.   Error bars are the standard deviation of the mean of up to $30$ shots.}
	\label{fig:fig3}
\end{figure}

\begin{figure}
	\centering
		\includegraphics[width=1.00\textwidth]{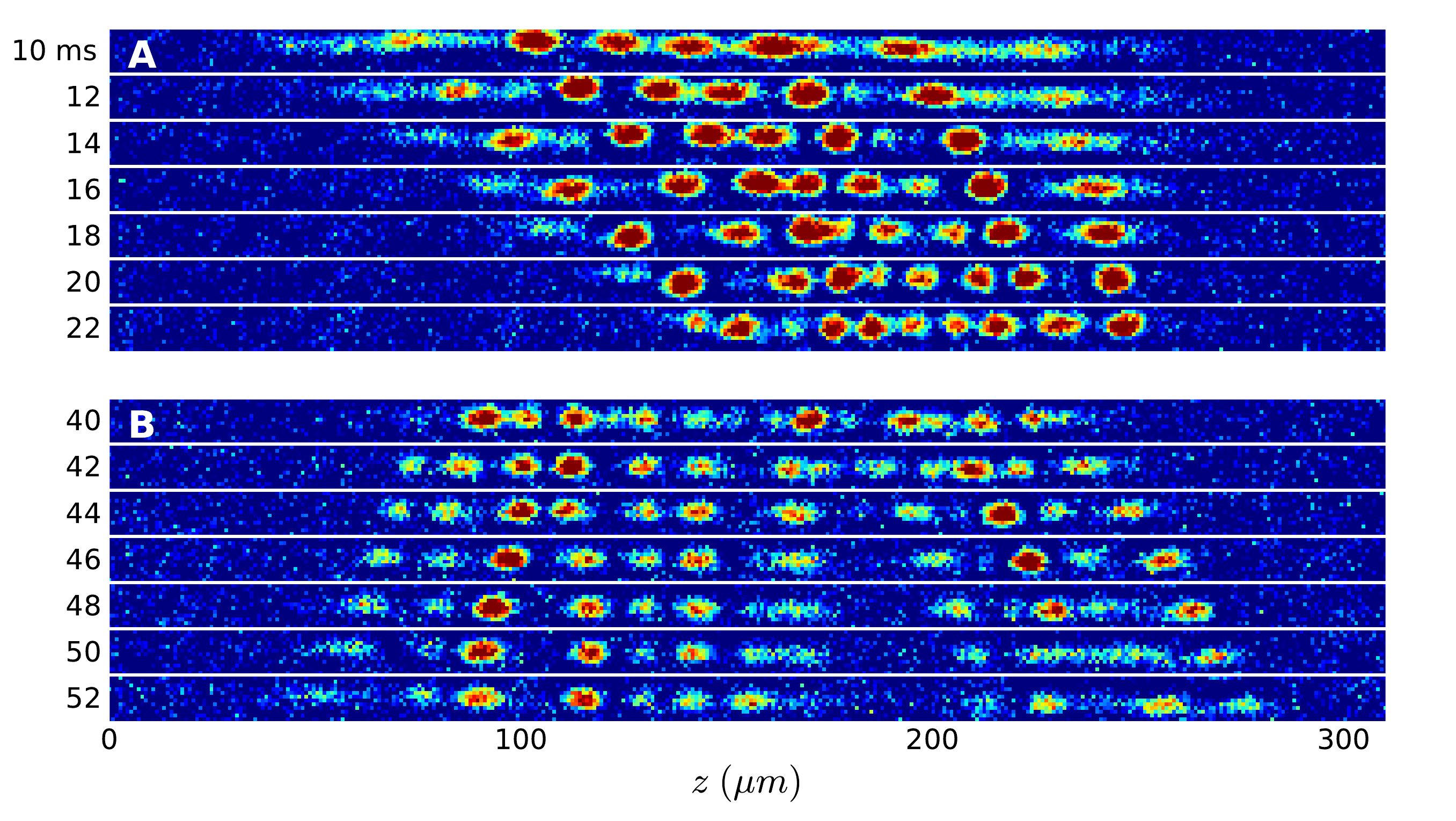}
	\caption*{\textbf{Fig. 4. Soliton train dynamics.} \textbf{(A)} Multiple images of the same soliton train, for $a_f=-0.18\ a_0$.  Beginning at $t_h=10\ \mathrm{ms}$, a new image was taken every $2\ \mathrm{ms}$.  We infer dominantly repulsive interactions although occasional attractive collisions occur between neighbors. The reduction in the overall size of the train is caused by a breathing mode excited by the quench, and a dipole oscillation is also evident.  \textbf{(B)} Similar to \textbf{A}, starting with $t_h=40\ \mathrm{ms}$.  The effects of the breathing mode in its expansion phase are evident.}
	\label{fig:fig4}
\end{figure}


\clearpage


\section*{Supplementary material}

# Supplementary Materials for
## Formation of matter-wave soliton trains by modulational instability

Jason H.V. Nguyen, De Luo, and Randall G. Hulet

Department of Physics and Astronomy, Rice University, Houston TX, 77005, USA

correspondence to: randy@rice.edu

**This PDF includes:**

- Materials and Methods
- Supplementary Text
- Fig. S1
- Fig. S2
- Fig. S3

# Materials and Methods

## Experimental Methods

The apparatus used here has been described previously (*26*). The harmonic trapping potential is derived from a laser beam operating at $1070$ nm, directed parallel to the magnetic field axis, and focused at the atoms to a $1/e^2$ radius of $32$ $\mu$m, giving us radial and axial trapping frequencies of $\omega_r/2\pi = 346$ Hz and $\omega_z/2\pi = 7.4$ Hz, respectively.

A pair of coils in the Helmholtz configuration is used to create a uniform magnetic field along the z-axis. The scattering length may be tuned across a broad region via the Feshbach resonance located at $738$ G (*38,39*). Initially, a BEC is formed at a field of $716$ G, corresponding to a scattering length of $a \approx 140$ $a_0$. The field is ramped in $250$ ms to a field of $565$ G where $a \approx 3$ $a_0$, corresponding to the initial conditions of all of our experiments. We find that a ramp time of $250$ ms is sufficiently slow that no breathing mode is excited. At this point, the field is quenched ($t_r = 1$ ms) to a final scattering length of $-2.5$ $a_0 < a_f < -0.18$ $a_0$. Within $\pm 4$ G of the zero-crossing at $543.6(1)$ G, the scattering length follows a linear relationship with a slope of $0.08(1)$ $a_0$/G (*39*).

To determine $N_a$, $N_s$, and $\Delta$, an *in situ* image of the atoms is taken using polarization phase contrast imaging (*40*), with the laser detuned from resonance by $20\Gamma$, where $\Gamma = 5.9$ MHz is the natural linewidth of $^7$Li. In order to examine the dynamics within a given experimental run, the imaging laser was detuned from resonance by $50\Gamma$, which afforded a good signal to noise ratio while keeping the atom loss below $2\%$ per image. $N_a$ is determined by integrating the column density image. $N_s$ is determined using a peak finding algorithm which subsequently fits the data to $N_s$ Gaussians, and from the fits we determine the separation between solitons, their $1/e$ axial length, and the total number of atoms in each soliton, which agrees to within $10\%$ of the number determined by doubly-integrating the image. For each data point in Figs. 2 and 3, we

take up to 30 averages, with error bars representing the standard deviation of the mean.

# Supplementary Text

## Calculation of $\eta$

To determine the effect of MI on the density of condensate, we integrate our images along the radial direction in order to obtain a line density, as shown in Fig. S1. The data is subsequently fit to a Thomas-Fermi distribution:

$$n(z) = \begin{cases} n_p \beta^2 + b, & \beta > 0 \\ b, & \beta < 0 \end{cases} \quad \text{(S1)}$$

where $\beta = 1 - \left(\frac{z-z_0}{z_{TF}}\right)^2$. We define a contrast parameter as:

$$\eta = \frac{\sqrt{\sum_i (n_e(i) - n_f(i))^2}}{n_p}, \quad \text{(S2)}$$

where $n_e(i)$ corresponds to the experimentally obtained line density at point $i$, and $n_f(i)$ is the obtained from the fit of the data at point $i$. We split the condensate into equal thirds, and calculate $\eta_L$ for the left–most region, $\eta_C$ for the central region, and $\eta_R$ for the right–most region. In Fig. S2 we plot $\eta$ vs. $t_h$ for $a_f = -0.18\, a_0$. We observe a rapid growth of $\eta_C$ as compared to $\eta_R$ and $\eta_L$, which we interpret as evidence that MI is predominantly seeded in the center of the distribution, and hence by noise, rather than by self-interference.

## Determination of the three-body decay rate

The decay in atom number resulting from three-body recombination can be analytically determined from:

$$\frac{1}{N_a}\frac{dN_a}{dt} = -\frac{L_3}{6}\langle n^2 \rangle, \quad \text{(S3)}$$

where $L_3$ is the three-body loss coefficient, and $\langle n^2 \rangle$ is the spatially-averaged second moment of the density. For simplicity, we determine the decay rate for a single soliton, and we consider

3

two separate cases. First, we assume that the density is well-described by the ground state of a harmonic oscillator, and thus the density is:

$$n(r,z) = \frac{N_a}{\pi^{3/2} a_r^2 a_z} e^{[-(r/a_r)^2]} e^{[-(z/a_z)^2]}, \tag{S4}$$

where $a_{r,z} = \sqrt{\hbar/(m\omega_{r,z})}$ is the harmonic oscillator length along the radial and axial directions, respectively. Using Eq. S4 we obtain:

$$\langle n^2 \rangle = \frac{N_a^2}{3\sqrt{3}\pi^3 a_r^4 a_z^2}, \tag{S5}$$

and using this we can solve Eq. S3, resulting in:

$$N_a(t_h) = \frac{N_0}{\left[\frac{L_3 N_0^2}{9\sqrt{3}\pi^3 a_r^4 a_z^2} t_h + 1\right]^{1/2}}, \tag{S6}$$

where $N_0$ is the initial atom number.

In obtaining Eq. S6 we have implicitly assumed that the length of the soliton is constant. In fact, the density may be more accurately described using the number-dependent wavefunction of a soliton:

$$n(r,z) = \frac{N_a}{4\pi a_r^2 \xi} e^{[-(r/a_r)^2]} \text{sech}^2\left(\frac{z}{2\xi}\right), \tag{S7}$$

where $\xi = \hbar^2/(m|g_{1D}|N_a)$ and $g_{1D} = 2\hbar\omega_r a$. Here, the axial length of the soliton changes as with changing $N_a$. In this case, the second moment of the density is given by:

$$\langle n^2 \rangle = \frac{m^2 g_{1D}^2}{90\pi^2 a_r^4 \hbar^4} N_a^4, \tag{S8}$$

and when this is used to solve Eq. S3 we obtain:

$$N_a(t_h) = \frac{N_0}{\left[\frac{m^2 g_{1D}^2 L_3 N_0^4}{135\pi^2 a_r^4 \hbar^4} t_h + 1\right]^{1/4}}, \tag{S9}$$

The power law behavior $N_a \propto t_h^{-1/2}$ for a fixed length, or $N_a \propto t_h^{-1/4}$ for a time-varying one brackets our fitted decay exponent of $\kappa = -0.35(1)$. We fit our data to Eqs. (S6) and

(S9) to determine $L_3$ (see Fig. S3). We assume that each soliton can be treated independently, and so we only fit the data for $a_f = -0.18\ a_0$ where $N_s \approx 11$ and $a_f = -0.42\ a_0$ where $N_s \approx 18$. For both scattering lengths, we only include data starting when we see a decrease in atom number, and up to $t_h = t_{br}/2$ (indicated by the dashed vertical line). We obtain a value of $L_3 \sim 10^{-26}\ \text{cm}^6/\text{s}$ for $a_f = -0.18\ a_0$ and $L_3 \sim 10^{-25}\ \text{cm}^6/\text{s}$ for $a_f = -0.42\ a_0$. In contrast, we find that when we slowly ramp the scattering length ($t_r = 500$ ms) to $a_f = -0.18\ a_0$, $N_s \approx 2$, and the atom number remains relatively flat, even though the initial number of atoms per soliton is similar to the quench data for $a_f = -0.18\ a_0$. From this, we are able to set an upper limit of $L_3 < 10^{-28}\ \text{cm}^6/\text{s}$. We conclude that our extracted values of $L_3$ arise from an enhancement of the density, either by the induced breathing mode, or partial collapses due to the quench.

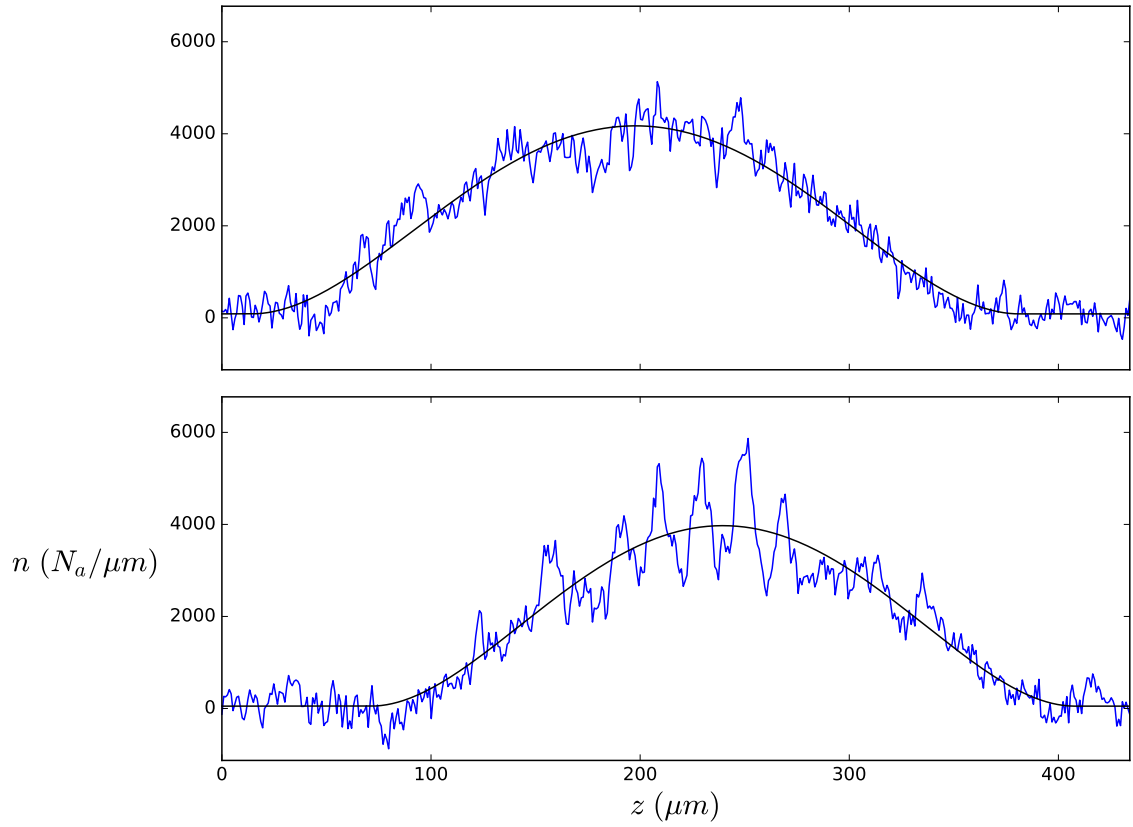

**Fig. S1. Representative Line density.** **(A)** Line density for $t_h = 2$ ms and $a_f = -0.18\, a_0$. The black solid line corresponds to fitting the data to Eq. 1. **(B)** Similar to **(A)** but with $t_h = 6$ ms.

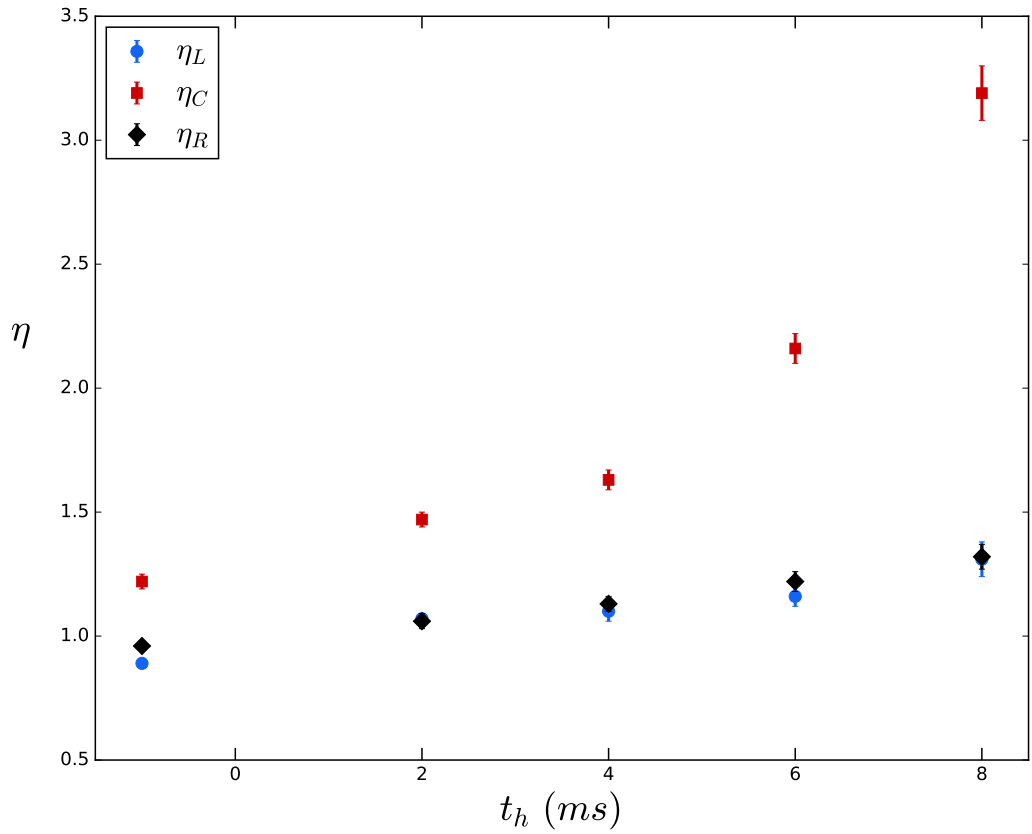

**Fig. S2. Onset of MI at early times.** $\eta$ vs. $t_h$ for $a_f = -0.18\, a_0$. We observe a rapid growth in $\eta_C$ as compared to $\eta_L$ and $\eta_R$. Each point corresponds to an average of up to 30 experimental runs.

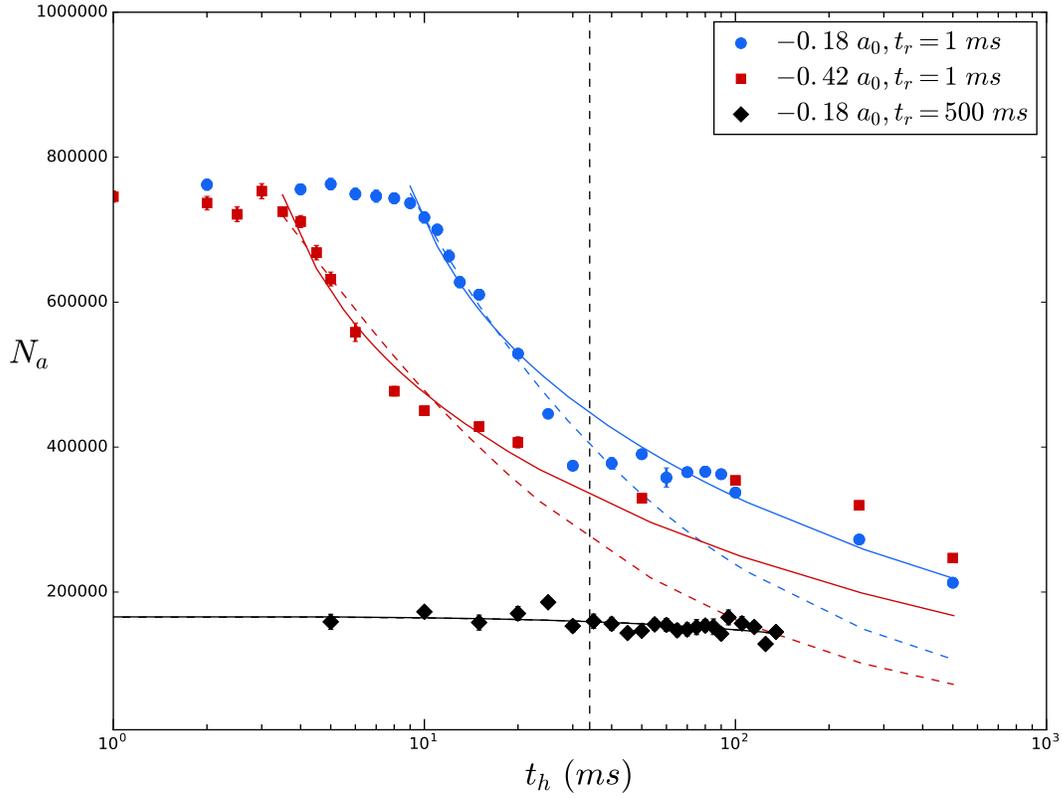

**Fig. S3. Fit to obtain three-body loss coefficient.** $N_a$ vs. $t_h$. The dashed lines corresponds to assuming a fixed soliton axial length while solid lines corresponds to allowing the soliton length to increase with decreasing number. For the slow ramp data (black) we are unable to detect atom loss for $t_h$ up to 100 ms.


\end{document}